\documentclass[journal]{IEEEtran}
\usepackage{rotating}% provides sidewaystable and sidewaysfigure
\usepackage{circuitikz}
\usepackage{tikz}
\usepackage{siunitx}
\usepackage{pgfplots}
\usepackage{multirow}
\usepackage{array}
\usepackage{threeparttable}
\usepackage{siunitx}
\usepackage{caption}
\usepackage{subcaption}
\usepackage{setspace}
\usepackage{amsmath}
\usetikzlibrary{decorations.markings}
\include{new_pmos}


\ifCLASSINFOpdf
\else
\fi
\begin{document}
\title{Design of Linear Passive Mixer-First Receivers for mmWave Digital Beamforming Arrays}

\author{Rawan~Al Kubaisy,~\IEEEmembership{Student~Member,~IEEE,}
        Sashank~Krishnamurthy,~\IEEEmembership{Member,~IEEE,}
        and~Ali~Niknejad,~\IEEEmembership{Fellow,~IEEE}% <-this % stops a space
\thanks{The authors are with the Berkeley Wireless Research Center, University of California, Berkeley, Berkeley,
CA, 94704 USA (e-mail:raalkubaisy@berkeley.edu).}% <-this % stops a space
}

\maketitle

\begin{abstract}
A 25-40GHz passive mixer-first receiver using a novel architecture for digital beamforming arrays is proposed. The architecture uses a novel technique for impedance matching using the on-resistance of the mixers in the receiver and matching networks. The small switch resistance of the mixers can be matched to the antenna using matching networks. Several matching networks are discussed, including a tunable matching network for wideband applications.  The design achieves a noise figure that is lower than 8dB, a conversion gain of 18dB, and an IIP3 of around +4dBm across the frequency range of 25-40GHz. A prototype chip is fabricated in 28nm bulk CMOS process.

\end{abstract}

% Note that keywords are not normally used for peer-review papers.
\begin{IEEEkeywords}
passive mixer-first receivers, wideband receivers, feedback linearization, overlapping LO, square LO, linear receivers, tunable matching network, charge sharing.
\end{IEEEkeywords}

% \fi
\IEEEpeerreviewmaketitle

\section{Introduction}
\label{chap: intro}
%\subsection{Motivation}
\IEEEPARstart{D}{igital} beamforming arrays are an enabler to the 5G revolution. The increased number of users necessitates switching to higher frequencies with a large number of frequency bands, this includes the 24-40GHZ band. The increasing number of bands in the mmWave spectrum mandates advanced circuit techniques to deal with the challenges associated with the high frequency and the large number of users. These challenges include linearity, interference mitigation, bandwidth, and other concerns. Due to the large number of users and antennas in MIMO systems, the use of digital beamforming arrays makes the linearity requirement more stringent but relaxes the noise figure requirement. This makes mixer-first receiver an attractive candidate for digital beamforming arrays because of their high linearity and their moderate noise figure.

%In MIMO systems, M>>K where M is the number of antennas and K is the number of users, this means that the number of antennas exceeds the number is users. 
%This Paper proposes circuit techniques for wideband linear mixer-first receiver. This thesis presents a novel mixer-first receiver architecture achieving noise figure values below 8dB, conversion gain of around 18dB, and IIP3 of 4dBm  for a frequency range of 25GHz to 40GHz. The design exploits feedback linearization techniques presented in \cite{JSSC_sashank} which is discussed in the following section.   
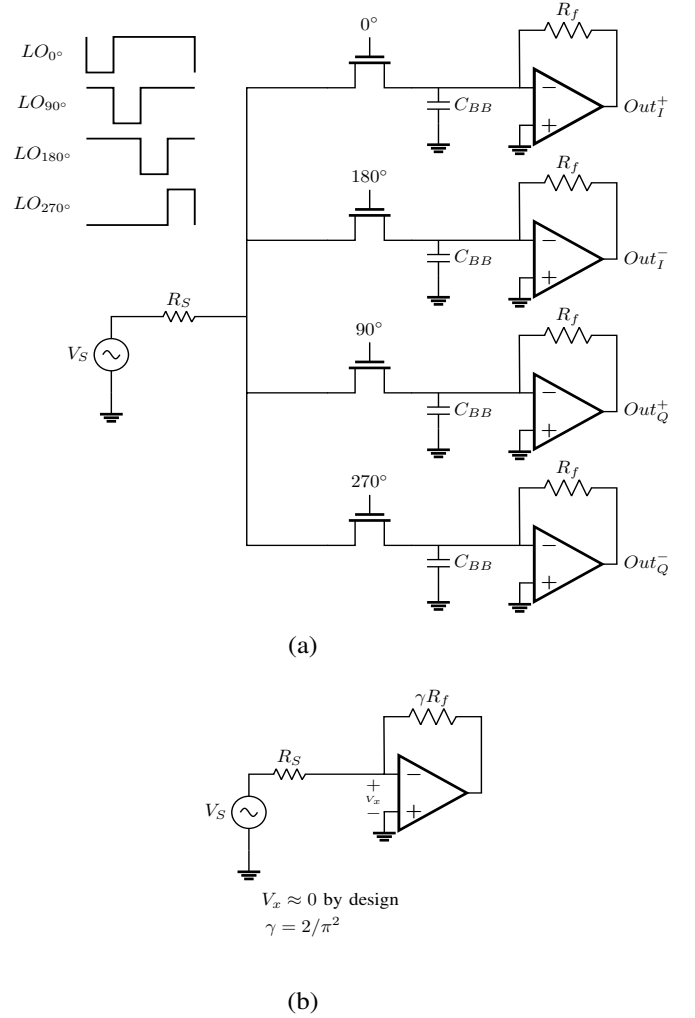
\begin{figure}
    \centering
     \resizebox{0.5\textwidth}{0.75\textwidth}{
\ctikzset{bipoles/thickness=1}

    \begin{circuitikz}[line width=0.75]

\draw(0,2) node[/tikz/circuitikz/bipoles/length=30pt,op amp, anchor=-,thick] (opamp) {}
 (opamp.out) to [short] ++(0,1.5) to [/tikz/circuitikz/bipoles/length=30pt,R,l_=$R_f$]  ++(-1.8,0) to (opamp.-)
(opamp.+) node[ground]{};
\draw(0,-1) node[/tikz/circuitikz/bipoles/length=30pt,op amp, anchor=-] (opamp1) {}
 (opamp1.out) to [short] ++(0,1.5) to [/tikz/circuitikz/bipoles/length=30pt,R,l_=$R_f$]  ++(-1.8,0) to (opamp1.-)
 (opamp1.+) node[ground]{};
\draw(0,-4) node[/tikz/circuitikz/bipoles/length=30pt,op amp, anchor=-] (opamp2) {}
 (opamp2.out) to [short] ++(0,1.5) to [/tikz/circuitikz/bipoles/length=30pt,R,l_=$R_f$]  ++(-1.8,0) to (opamp2.-)
(opamp2.+) node[ground]{};
\draw(0,-7) node[/tikz/circuitikz/bipoles/length=30pt,op amp, anchor=-] (opamp3) {}
 (opamp3.out) to [short] ++(0,1.5) to [/tikz/circuitikz/bipoles/length=30pt,R,l_=$R_f$]  ++(-1.8,0) to (opamp3.-)
 (opamp3.+) node[ground]{};

\draw (opamp.out) node[right]{$Out_I^+$};
\draw (opamp1.out) node[right]{$Out_I^-$};
\draw (opamp2.out) node[right]{$Out_Q^+$};
\draw (opamp3.out) node[right]{$Out_Q^-$};

\draw (opamp.-) to [short]  ++(-2,0) node[nmos,anchor=S,rotate=-90,yscale=-1](nmos_0){};
\draw (opamp1.-) to [short] ++(-2,0) node[nmos,anchor=S,rotate=-90,yscale=-1](nmos_180){};

\draw (opamp2.-)  to [short]  ++(-2,0) node[nmos,anchor=S,rotate=-90,yscale=-1](nmos_90){};
\draw (opamp3.-)  to [short]  ++(-2,0) node[nmos,anchor=S,rotate=-90,yscale=-1](nmos_270){};

\draw (nmos_270.G) to [open] ++(0,-0.1) node[label=$270^\circ$]{};
\draw (nmos_180.G) to [open] ++(0,-0.1) node[label=$180^\circ$]{};
\draw (nmos_0.G) to [open] ++(0,-0.1) node[label=$0^\circ$]{};
\draw (nmos_90.G) to [open] ++(0,-0.1) node[label=$90^\circ$]{};

\draw (nmos_0.D) to [/tikz/circuitikz/bipoles/length=20pt,short] ++(-1.5,0) to [short] ++(0,-9) to [short] ++(1.5,0) to [open] ++(0,3) to [/tikz/circuitikz/bipoles/length=20pt,short] ++(-1.5,0) to [open] ++(0,3) to [short] ++(1.5,0);

\draw (nmos_90.D)++(-1.5,1.5) to [/tikz/circuitikz/bipoles/length=20pt,R,l_=$R_S$] ++(-2.5,0) to [/tikz/circuitikz/bipoles/length=30pt,american,sV,l_=$V_S$] ++(0,-1.5)  node[ground]{};

\draw(opamp.-)++(-1.5,0) to [/tikz/circuitikz/bipoles/length=20pt,C=$C_{BB}$] ++(0,-0.7) node[ground]{};
\draw(opamp1.-)++(-1.5,0) to [/tikz/circuitikz/bipoles/length=20pt,C=$C_{BB}$] ++(0,-0.7) node[ground]{};
\draw(opamp2.-)++(-1.5,0) to [/tikz/circuitikz/bipoles/length=20pt,C=$C_{BB}$] ++(0,-0.7) node[ground]{};
\draw(opamp3.-)++(-1.5,0) to [/tikz/circuitikz/bipoles/length=20pt,C=$C_{BB}$] ++(0,-0.7) node[ground]{};

        \draw(-2.5,-11.5) node[/tikz/circuitikz/bipoles/length=30pt,op amp, anchor=-] (opamp4) {}
 (opamp4.out) to [short] ++(0,1.5) to [/tikz/circuitikz/bipoles/length=30pt,R,l_=$\gamma R_f$]  ++(-1.8,0) to (opamp4.-) to [short] ++(-1,0) to [/tikz/circuitikz/bipoles/length=20pt,R,l_=$R_S$] ++(-1.5,0) to [/tikz/circuitikz/bipoles/length=30pt,sV,l_=$V_S$] ++(0,-1.5) node[ground]{}
 (opamp4.+) node[ground]{};
      \draw [line width=1pt]  (-8,3) node[anchor=south west]{} to [short] ++(0,-0.7) to [short] ++(0.5,0) to [short] ++(0,0.7) to [short] ++(1.5,0) to [short] ++(0,-0.7);
    \draw [line width=1pt]  (-8,2) node[anchor= west]{} to [short] ++(0.5,0) to [short] ++(0,-0.70) to [short] ++(0.5,0) to [short] ++(0,0.7) to [short] ++(1,0);
    \draw [line width=1pt]  (-8,1) to [short] ++(1,0) to [short] ++(0,-0.7) to [short] ++(0.5,0) to [short] ++(0,0.7) to [short] ++(0.5,0);
    \draw [line width=1pt] (-8,-0.7)  to [short] ++(1.5,0) to [short] ++(0,0.7) to [short] ++(0.5,0) to [short] ++(0,-0.7);
    
    \draw(-8.8,2.7) node[]{$LO_{0^\circ}$};
    \draw(-8.8,1.7) node[]{$LO_{90^\circ}$};
    \draw(-8.8,0.7) node[]{$LO_{180^\circ}$};
    \draw(-8.8,-0.3) node[]{$LO_{270^\circ}$};
    
    \draw (opamp4.-)++(-0.2,-.1) to [open,american,v=\tiny$V_x$] ++(0,-0.8); 
    \draw(-3.5,-14) node[]{$V_x\approx0$ by design};
    \draw(-4,-14.5) node[]{$\gamma=2/\pi^2$};

    \draw(-4,-9) node[]{\Large(a)};
    \draw(-4,-16) node[]{\Large(b)};

\end{circuitikz}}

 \caption{(a) Conventional 4-phase passive mixer-first receiver and non-overlapping LO waveforms. (b) Its LTI equivalent.}
    \label{fig:conventionalnpath}
\end{figure}

%\subsection{Prior Work on High-linearity mixer-first receiver}
Fig. \ref{fig:conventionalnpath} shows a conventional 4-phase passive mixer-first receiver driven by non-overlapping LO waveform. Traditionally, matching is done using the transparency property of N-path filters. Instead of using a shunt resistor, which will add a 3dB penalty to the noise figure, matching is done using miller's 's effect with the feedback resistor of the baseband amplifier. Using miller's 's  effect, the impedance looking into the receiver from the antenna is equal to:
%each baseband amplifier is equal to 
\begin{equation}
    Z_{in}=\frac{\gamma R_f}{(1+A)}
\end{equation}
%The baseband impedance is scaled by the factor $\gamma=2/\pi^2$. 

A high linearity mixer-first receiver for digital mmWave beamforming arrays was proposed in \cite{JSSC_sashank}. Fig. \ref{fig:HighLinRx}.a shows the schematic of the highly linear receiver. The work achieves in-band IIP3 that are 16dB higher than the state-of-the-art passive mixer-first receivers. Instead of using miller's  effect of the feedback resistor of the baseband amplifier for matching, matching is done with a 50$\Omega$ physical resistor. The design uses feedback linearization to improve the linearity of the receiver. 

Feedback linearization is achieved by designing a baseband amplifier with a large open loop gain, making the voltage at the input of the baseband amplifier a virtual ground. The large loop gain of the baseband amplifier would make the impedance looking into the baseband amplifier small and reducing the swing at the input of the amplifier. The small input swing would translate to less distortion caused by the baseband amplifier. Fig. \ref{fig:HighLinRx}.b shows an illustration of that where $V_x$ is chosen to be approximately zero by design. 

The noise figure of the design ranges from 12.5dB to 15.7dB for $f_{RF}$ that ranges from 10GHZ to 30GHz. The noise figure is high due to several factors. The use of the 50$\Omega$ resistors adds a 3.1dB penalty to the noise figure. Additionally, charge sharing plays a large role in degrading the noise figure of receivers with overlapping local LO waveform. Synthesizing 25\% duty-cycle LO Waveform at microwave and mmWave frequencies is challenging. Hence, a 50\% duty-cycle LO is used in this design instead. The overlap of LO waveform results in charge sharing between the I and Q paths which will degrade the noise figure. The authors in \cite{JSSC_sashank} proposed using 50$\Omega$ resistors in all of the mixers' paths, instead of one 50$\Omega$ resistor to reduce the charge sharing. Although the addition of the resistors on all of the paths helped reduce charge sharing, the resistors would only reduce charge sharing current and not filter it out. On the other hand, the addition of resistors on all paths increased the parasitic capacitance leading to more loss in the signal path, and a further degradation in the noise figure. 

This paper is organized as follows. Section \ref{chap:2} explores a narrowband mixer-first receiver design using a modified version of the Wilkinson divider. Section \ref{chap:3} explores the use of L-matching networks in the receiver including a tunable matching network for wideband applications. Section \ref{chap:4} discuss the circuit implementation of the wideband receiver discussed in section \ref{chap:3} and the post-layout simulations. Section \ref{chap:5} compares this work against other mixer-first receivers and provides the takeaway and possible way to improve the design. 

\begin{figure}
    \centering
     \resizebox{0.5\textwidth}{0.75\textwidth}{
\ctikzset{bipoles/thickness=1}

    \begin{circuitikz}[line width=0.75]

\draw(0,2) node[/tikz/circuitikz/bipoles/length=30pt,op amp, anchor=-,thick] (opamp) {}
 (opamp.out) to [short] ++(0,1.5) to [/tikz/circuitikz/bipoles/length=30pt,R,l_=$R_f$]  ++(-1.8,0) to (opamp.-)
(opamp.+) node[ground]{};
\draw(0,-1) node[/tikz/circuitikz/bipoles/length=30pt,op amp, anchor=-] (opamp1) {}
 (opamp1.out) to [short] ++(0,1.5) to [/tikz/circuitikz/bipoles/length=30pt,R,l_=$R_f$]  ++(-1.8,0) to (opamp1.-)
 (opamp1.+) node[ground]{};
\draw(0,-4) node[/tikz/circuitikz/bipoles/length=30pt,op amp, anchor=-] (opamp2) {}
 (opamp2.out) to [short] ++(0,1.5) to [/tikz/circuitikz/bipoles/length=30pt,R,l_=$R_f$]  ++(-1.8,0) to (opamp2.-)
(opamp2.+) node[ground]{};
\draw(0,-7) node[/tikz/circuitikz/bipoles/length=30pt,op amp, anchor=-] (opamp3) {}
 (opamp3.out) to [short] ++(0,1.5) to [/tikz/circuitikz/bipoles/length=30pt,R,l_=$R_f$]  ++(-1.8,0) to (opamp3.-)
 (opamp3.+) node[ground]{};

\draw (opamp.out) node[right]{$Out_I^+$};
\draw (opamp1.out) node[right]{$Out_I^-$};
\draw (opamp2.out) node[right]{$Out_Q^+$};
\draw (opamp3.out) node[right]{$Out_Q^-$};

\draw (opamp.-) to [short]  ++(-2,0) node[nmos,anchor=S,rotate=-90,yscale=-1](nmos_0){};
\draw (opamp1.-) to [short] ++(-2,0) node[nmos,anchor=S,rotate=-90,yscale=-1](nmos_180){};

\draw (opamp2.-)  to [short]  ++(-2,0) node[nmos,anchor=S,rotate=-90,yscale=-1](nmos_90){};
\draw (opamp3.-)  to [short]  ++(-2,0) node[nmos,anchor=S,rotate=-90,yscale=-1](nmos_270){};

\draw (nmos_270.G) to [open] ++(0,-0.1) node[label=$270^\circ$]{};
\draw (nmos_180.G) to [open] ++(0,-0.1) node[label=$180^\circ$]{};
\draw (nmos_0.G) to [open] ++(0,-0.1) node[label=$0^\circ$]{};
\draw (nmos_90.G) to [open] ++(0,-0.1) node[label=$90^\circ$]{};

\draw (nmos_0.D) to [/tikz/circuitikz/bipoles/length=20pt,R,l_=$R_S'$] ++(-1.5,0) to [short] ++(0,-9) to [/tikz/circuitikz/bipoles/length=20pt,R=$R_S'$] ++(1.5,0) to [open] ++(0,3) to [/tikz/circuitikz/bipoles/length=20pt,R,l_=$R_S'$] ++(-1.5,0) to [open] ++(0,3) to [/tikz/circuitikz/bipoles/length=20pt,R,l=$R_S'$] ++(1.5,0);

\draw (nmos_90.D)++(-1.5,1.5) to [/tikz/circuitikz/bipoles/length=20pt,R,l_=$R_S$] ++(-2.5,0) to [/tikz/circuitikz/bipoles/length=30pt,american,sV,l_=$V_S$] ++(0,-1.5)  node[ground]{};

\draw(opamp.-)++(-1.5,0) to [/tikz/circuitikz/bipoles/length=20pt,C=$C_{BB}$] ++(0,-0.7) node[ground]{};
\draw(opamp1.-)++(-1.5,0) to [/tikz/circuitikz/bipoles/length=20pt,C=$C_{BB}$] ++(0,-0.7) node[ground]{};
\draw(opamp2.-)++(-1.5,0) to [/tikz/circuitikz/bipoles/length=20pt,C=$C_{BB}$] ++(0,-0.7) node[ground]{};
\draw(opamp3.-)++(-1.5,0) to [/tikz/circuitikz/bipoles/length=20pt,C=$C_{BB}$] ++(0,-0.7) node[ground]{};

    \draw [line width=1pt]  (-8,3) node[anchor=south west]{} to [short] ++(0,-0.7) to [short] ++(1,0) to [short] ++(0,0.7) to [short] ++(1,0) to [short] ++(0,-0.7);
    \draw [line width=1pt]  (-8,2) node[anchor= west]{} to [short] ++(0.5,0) to [short] ++(0,-0.70) to [short] ++(1,0) to [short] ++(0,0.7) to [short] ++(0.5,0);
    \draw [line width=1pt]  (-8,0.3)  to [short] ++(0,0.7) node[left]{} to [short] ++(1,0) to [short] ++(0,-0.7) to [short] ++(1,0) to [short] ++(0,0.7);
    \draw [line width=1pt] (-8,-0.7)  to [short] ++(0.5,0) node[left]{} to [short] ++(0,0.7) to [short] ++(1,0) to [short] ++(0,-0.7) to [short] ++(0.5,0);
    
    \draw(-8.8,2.7) node[]{$LO_{0^\circ}$};
    \draw(-8.8,1.7) node[]{$LO_{90^\circ}$};
    \draw(-8.8,0.7) node[]{$LO_{180^\circ}$};
    \draw(-8.8,-0.3) node[]{$LO_{270^\circ}$};
    
        \draw(-2.5,-11.5) node[/tikz/circuitikz/bipoles/length=30pt,op amp, anchor=-] (opamp4) {}
 (opamp4.out) to [short] ++(0,1.5) to [/tikz/circuitikz/bipoles/length=30pt,R,l_=$\gamma R_f$]  ++(-1.8,0) to (opamp4.-) to [/tikz/circuitikz/bipoles/length=20pt,R,l_=$R_{S}'$] ++(-1,0) to [/tikz/circuitikz/bipoles/length=20pt,R,l_=$R_S$] ++(-1.5,0) to [/tikz/circuitikz/bipoles/length=30pt,sV,l_=$V_S$] ++(0,-1.5) node[ground]{}
 (opamp4.+) node[ground]{};
  \draw (opamp4.-)++(-0.2,-.1) to [open,american,v=\tiny$V_x$] ++(0,-0.8); 
    \draw(-3.5,-14) node[]{$V_x\approx0$ by design};
    \draw(-4,-14.5) node[]{$\gamma=2/\pi^2$};

    \draw(-4,-9) node[]{\Large(a)};
    \draw(-4,-16) node[]{\Large(b)};

\end{circuitikz}}

 \caption{High-linearity mixer-first receiver (a) full schematic and the overlapping LO waveforms. (b) The LTI equivalent.}
    \label{fig:HighLinRx}
\end{figure}
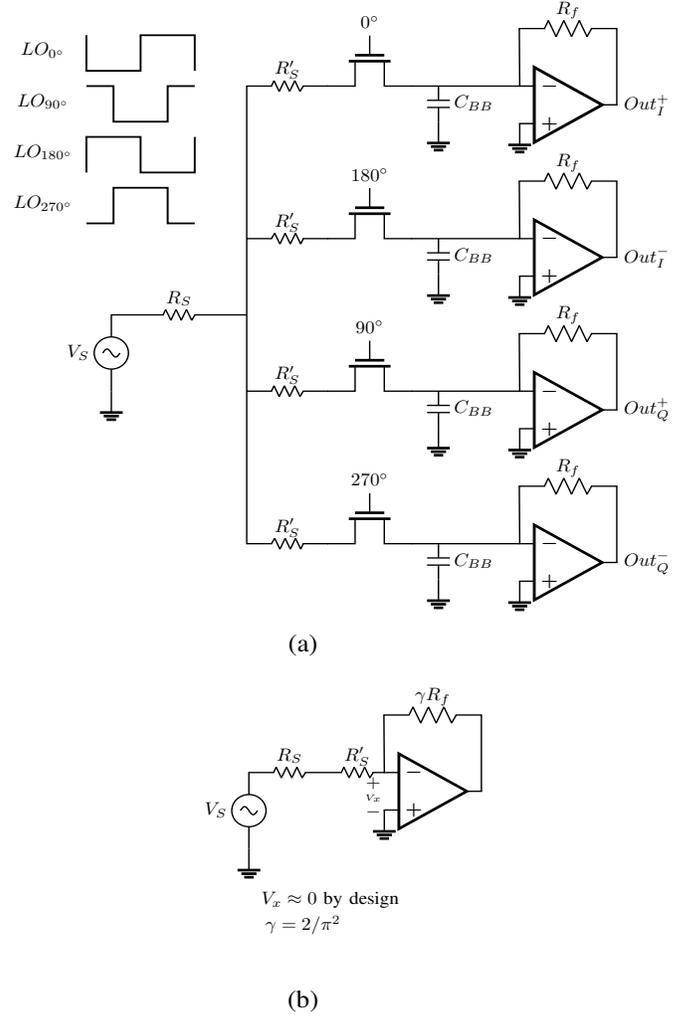

\begin{figure}
    \centering
         \resizebox{0.5\textwidth}{0.55\textwidth}{
%\tikzset{bipoles/thickness=1}
%\begin{circuitikz}[line width=1pt]
\begin{tikzpicture} 
        %Environment Config
        %Style Variable
       [ text pos/.store in=\tpos,text pos=0.5,
        text anchor/.store in=\tanchor,text anchor={north:12pt},
        Tline/.style={%Style for the voltage reference
            draw,
            postaction={decorate,decoration={markings,mark=at position 10pt with {\coordinate (a) at (90:3.5pt);}}},
            postaction={decorate,decoration={markings,mark=at position \tpos with {\node at (\tanchor){\small #1};}}},
            postaction={decorate,decoration={markings,mark=at position \pgfdecoratedpathlength-10pt with {\coordinate (b) at (-90:3.5pt);\draw[fill=black!40](a) rectangle (b);}}}}]

\draw(0,2) node[/tikz/circuitikz/bipoles/length=30pt,op amp, anchor=-] (opamp) {}
 (opamp.out) to [short,thick] ++(0,1.5) to [/tikz/circuitikz/bipoles/length=30pt,R,l_=$R_f$,thin]  ++(-1.8,0) to (opamp.-)
(opamp.+) node[ground]{};
\draw(0,-1) node[/tikz/circuitikz/bipoles/length=30pt,op amp, anchor=-] (opamp1) {}
 (opamp1.out) to [short,thick] ++(0,1.5) to [/tikz/circuitikz/bipoles/length=30pt,R,l_=$R_f$,thin]  ++(-1.8,0) to (opamp1.-)
 (opamp1.+) node[ground]{};
\draw(0,-4) node[/tikz/circuitikz/bipoles/length=30pt,op amp, anchor=-] (opamp2) {}
 (opamp2.out) to [short] ++(0,1.5) to [/tikz/circuitikz/bipoles/length=30pt,R,l_=$R_f$]  ++(-1.8,0) to (opamp2.-)
(opamp2.+) node[ground]{};
\draw(0,-7) node[/tikz/circuitikz/bipoles/length=30pt,op amp, anchor=-] (opamp3) {}
 (opamp3.out) to [short] ++(0,1.5) to [/tikz/circuitikz/bipoles/length=30pt,R,l_=$R_f$]  ++(-1.8,0) to (opamp3.-)
 (opamp3.+) node[ground]{};

\draw (opamp.out) node[right]{$Out_I^+$};
\draw (opamp1.out) node[right]{$Out_I^-$};
\draw (opamp2.out) node[right]{$Out_Q^+$};
\draw (opamp3.out) node[right]{$Out_Q^-$};

\draw (opamp.-) to [short]  ++(-2,0) node[nmos,anchor=S,rotate=-90,yscale=-1](nmos_0){};
\draw (opamp1.-) to [short] ++(-2,0) node[nmos,anchor=S,rotate=-90,yscale=-1](nmos_180){};

\draw (opamp2.-)  to [short]  ++(-2,0) node[nmos,anchor=S,rotate=-90,yscale=-1](nmos_90){};
\draw (opamp3.-)  to [short]  ++(-2,0) node[nmos,anchor=S,rotate=-90,yscale=-1](nmos_270){};

\draw (nmos_270.G) to [open] ++(0,-0.1) node[label=$270^\circ$]{} ;
\draw (nmos_180.G) to [open] ++(0,-0.1) node[label=$180^\circ$]{} ;
\draw (nmos_0.G) to [open] ++(0,-0.1) node[label=$0^\circ$]{} ;
\draw (nmos_90.G) to [open] ++(0,-0.1) node[label=$90^\circ$]{} ;

\draw (nmos_90.D) to [short] (nmos_270.D);
\draw (nmos_0.D) to [short] (nmos_180.D);

\draw [Tline] (nmos_0.D)++(-1.5,-1.5)-- ++(0,-3);
\draw [Tline] (nmos_90.D)++(-1.5,-1.5)-- ++(0,3);
\draw (nmos_90.D)++(0,-1.5) to [short] ++(-1.5,0);
\draw (nmos_0.D)++(0,-1.5) to [short] ++(-1.5,0);

\draw (-5.05,-2.5) to [/tikz/circuitikz/bipoles/length=20pt,R,l_=$R_S$] ++(-1.5,0) to [/tikz/circuitikz/bipoles/length=30pt,american,sV,l_=$V_S$] ++(0,-1.5)  node[ground]{};;
%=$\lambda/4$,text anchor=-90:12pt
%\draw [Tline] (nmos_270.D) ++(-1,1.5) --++(0,2);
\draw (-4.5,-1.75) node[label=$\lambda/4$]{};
%\draw [Tline] (nmos_90.D) -- ++(-1.5,1.5);
\draw (-4.5,-4.2) node[label=$\lambda/4$]{};

\draw(opamp.-)++(-1.5,0) to [/tikz/circuitikz/bipoles/length=20pt,C=$C_{BB}$] ++(0,-0.7) node[ground]{};
\draw(opamp1.-)++(-1.5,0) to [/tikz/circuitikz/bipoles/length=20pt,C=$C_{BB}$] ++(0,-0.7) node[ground]{};
\draw(opamp2.-)++(-1.5,0) to [/tikz/circuitikz/bipoles/length=20pt,C=$C_{BB}$] ++(0,-0.7) node[ground]{};
\draw(opamp3.-)++(-1.5,0) to [/tikz/circuitikz/bipoles/length=20pt,C=$C_{BB}$] ++(0,-0.7) node[ground]{};
\end{tikzpicture}}
    \caption{Mixer-first Receiver with Modified Wilkinson Divider}
    \label{fig:WilkinsonRx2}
\end{figure}
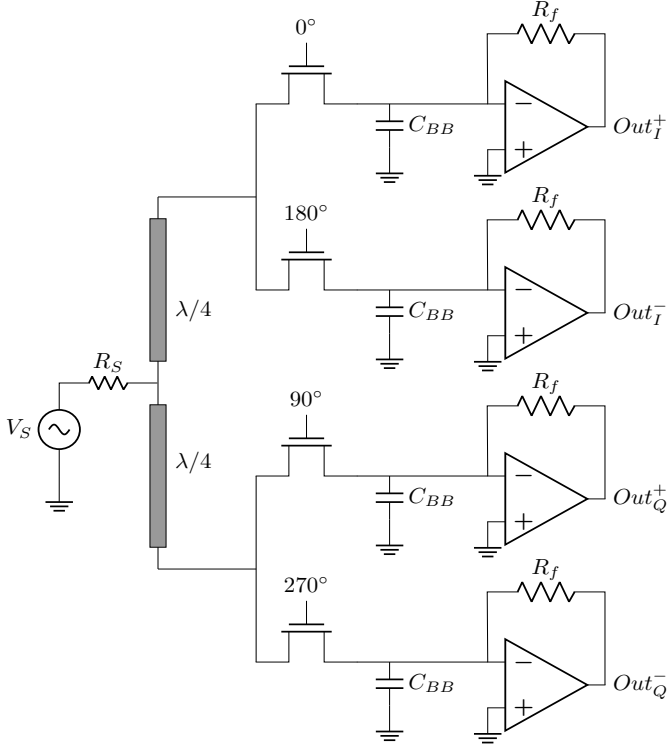

\section{Mixer-First Receiver with Modified Wilkinson Divider}
\label{chap:2}
In an effort to improve the noise figure of the linear receiver proposed in \cite{JSSC_sashank}, this paper proposes eliminating the 50$\Omega$ physical resistor and using the on-resistance of the mixers for matching. With the help of quarter-wavelength transmission lines, the small value for on-resistance of the mixers' can be transformed to a larger value to match the antenna. Fig. \ref{fig:WilkinsonRx2} shows the schematic of this design. The quarter-wavelength transmission lines structure is similar to Wilkinson Divider but without the bridge resistor.

Impedance matching can be achieved by choosing the width of the transmission line such that its characteristic impedance is
  
\begin{equation} \label{eq:2.1}
Z_0=\sqrt{2R_{sw}R_S}
\end{equation}
this makes the impedance looking into one transmission line, which is either the I or the Q paths, equal to
\begin{equation} \label{eq:2.2}
Z_{Tline1}=\frac{Z_0^2}{R_{sw}}=\frac{2R_{sw}R_{s}}{R_{sw}}=2R_{S}
\end{equation}
And the input impedance looking into the receiver is
\begin{equation} \label{eq:2.3}
Z_{in}=2R_{S}||2R_{S}=R_S
\end{equation}

%The second proposed design uses a modified version of a Wilkinson divider between the antenna and the mixers. Common-mode half circuit analysis in figure \ref{fig:WilkinsonMatching} shows that the bridge resistor contribute to the input matching because the bridge resistor is open in odd mode half circuit. The main purpose of the resistor is to provide isolation between the $Port_2$ and $Port_3$. The discussion at the end of section \ref{sec: limit} showed that the bridge resistor doesn't help with suppressing the charge sharing current, and it might be increasing the charge sharing between the I and the Q paths and impacting the noise figure. Hence, the bridge resistor can be omitted from the design. The rest of the design is not changed from the previous architecture. The quarter-wavelength transmission lines are still used to transform the on-resistance of mixers to match the antenna. Similar to the previous architecture, the on-resistance of the mixers is used for matching with the quarter-wavelength transformer included.   
%This design uses feedback linearization proposed in the previous section to mitigate the effect of the baseband distortion on receiver's overall linearity.

%The simulation setup is similar to setup of the simulation with the full Wilkinson Divider. The only difference is that the bridge resistor is removed from the schematics. 

The addition of the quarter-wavelength transmission lines will provide isolation between the I and the Q paths of the receiver. This means that the charge sharing current that is caused by the overlapping LO waveform will be reduced. To achieve the same linearity measurements as in \cite{JSSC_sashank}, this design uses feedback linearization proposed in \cite{JSSC_sashank} and discussed in section \ref{chap: intro} to mitigate the effect of the baseband distortion on the receiver's overall linearity. Feedback linearization means choosing a large open loop gain for the baseband amplifier which would result in a smaller voltage swing at the input of the baseband amplifier.

\begin{figure}
    \centering

\begin{tikzpicture}
\begin{axis}[
    width=0.5\textwidth,
    height=0.35\textwidth,
    xlabel={Frequency [GHz]},
    ylabel={Noise Figure [dB]},
    xmin=10, xmax=50,
    ymin=5, ymax=16,
    xtick={10,15,20,25,30,35,40,45,50},
    ytick={5,6,7,8,9,10,11,12,13,14,15,16},
    legend pos=north west,
    ymajorgrids=true,
    xmajorgrids=true,
    x label style={at={(axis description cs:0.5,0.0)}},
    y label style={at={(axis description cs:0.07,0.5)}},
]
\addplot[
    color=black, mark=o, mark size = {2},  thick
    ]
    coordinates {
    (10,10)(15,10.54)(20,11.16)(25,11.81)(30,12.48)(35,13.14)(40,13.8)(45,14.43)(50,15)
    };
    
    \addlegendentry{Rx W/ 50$\Omega$ resistor}
\addplot[
    color=blue, mark=square, mark size = {2}, thick]
    coordinates {
    (10,8.74)(15,7.3)(20,6.69)(25,6.31)(30,6.11)(35,6.81)(40,9.1)(45,12.36)(50,15.68)
    };
    \addlegendentry{Rx W/ Tlines}

    \node (source) at (axis cs:30,6.21){};
    \node (destination) at (axis cs:30,12.38){};
    \draw [thick, <->] (source) -- (destination);
    \node (label) at (axis cs:33.5,9.5){6.37dB};
    
\end{axis}
\end{tikzpicture}
    \caption{Noise figure of receiver with a 50$\Omega$ resistor and receiver with a $\lambda/4$ transmission line designed for 30GHz}
    \label{fig:WilkinsonRx2-30G}
\end{figure}
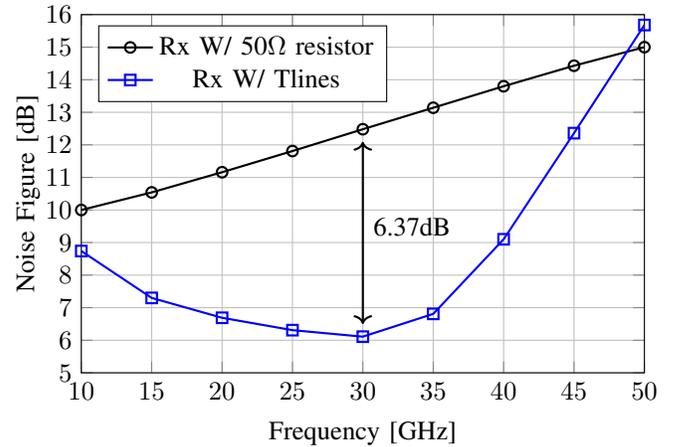

Fig. \ref{fig:WilkinsonRX2Results} shows the simulation results of the design. There is a 4.7dB improvement in the noise figure compared to the 50$\Omega$ resistors design proposed in \cite{JSSC_sashank} at 20GHz. Fig. \ref{fig:WilkinsonRx2-30G} also shows the simulated noise figure values for the same design with transmission lines designed for 30GHz with a 6.1dB improvement in noise figure. The higher frequency means a shorter length for the quarter-wavelength transmission lines, less loss in the signal path, and an improved noise figure. 

Using transmission lines in a mixer-first receivers significantly improved the noise figure of the linear receiver proposed in \cite{JSSC_sashank}. But the use of transmission lines is limited due to several factors. At 20GHz, the required length of a quarter-wavelength transmission line is about 2mm. This means that the design will consume a large area and will not be practical. Additionally, this receiver design can only be used for narrowband applications since matching will only be achieved at a single frequency. 
\section{Mixer-first Receiver with Tunable Matching Network}
\label{chap:3}

This section discusses a receiver design to overcome the area and the bandwidth limitation imposed by transmission lines in the design proposed in section \ref{chap:2}. The design still uses the on-resistance of the mixers' for matching, but the quarter-wavelength transmission lines are replaced with a passive L-matching network. The matching network will behave as an artificial quarter-wavelength transmission lines and it will transform the on-resistance of the mixers to match the 50$\Omega$ resistance of the antenna. Fig. \ref{fig:RX_MN1} shows a schematic of the receiver with the passive matching network. The matching network is used on both the I and the Q paths, and it behaves as a low-pass filter to isolate the I and Q paths by filtering the charge sharing current between the two paths due to the overlapping LO waveform. It consists of a shunt cap at the input and a series inductor. This configuration is chosen as opposed to the capacitor in series and a shunt inductor because the series inductor would provide filtering to the charge sharing current, which is a technique proposed in \cite{Molnar_indu}. The use of an L-matching network is an improvement from the previous design because a \SI{2}{\milli\meter} transmission line is no longer needed, which puts less constraint on the area. 
\begin{figure}
    \centering

\begin{tikzpicture}
\begin{axis}[
    width=0.5\textwidth,
    height=0.35\textwidth,
    xlabel={Frequency [GHz]},
    ylabel={Noise Figure [dB]},
    xmin=10, xmax=35,
    ymin=5, ymax=15,
    xtick={10,15,20,25,30,35},
    ytick={5,6,7,8,9,10,11,12,13,14,15},
    legend pos=north west,
    %ymajorgrids=true,
    %xmajorgrids=true,
    grid=both,
    x label style={at={(axis description cs:0.5,0.0)}},
    y label style={at={(axis description cs:0.07,0.5)}},
]

\addplot[color=black, mark=o, mark size = {2}, thick]
    coordinates {
    (10,10)(15,10.54)(20,11.16)(25,11.81)(30,12.48)(35,13.14)
    };
    
    \addlegendentry{Rx W/ 50$\Omega$ resistor}
\addplot[color=blue, mark=square, mark size = {2}, thick,
    ]
    coordinates {
    (10,7.34)(15,6.56)(20,6.41)(25,7.35)(30,10.38)(35,14.5)
    };
    \addlegendentry{Rx W/ Tlines}

    \node (source) at (axis cs:20,6.51){};
    \node (destination) at (axis cs:20,11.06){};
    \draw [thick, <->] (source) -- (destination);
    \node (label) at (axis cs:22,9.5){4.75dB};
\end{axis}
\end{tikzpicture}
    \caption{Noise figure of receiver with a 50$\Omega$ resistor and receiver with a $\lambda/4$ transmission line designed for 20GHz}
    \label{fig:WilkinsonRX2Results}
\end{figure}
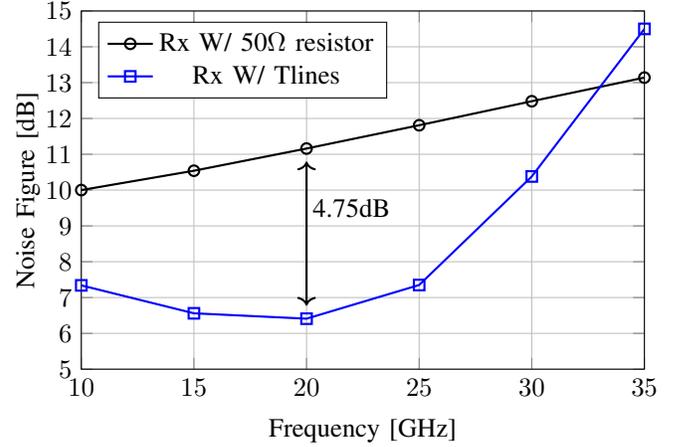

Fig. \ref{fig:RX_MN} shows the simulation results of the three receiver designs. The improvement in noise figure with the use of the passive matching network is similar to the improvement seen with the transmission lines design. The receiver design with the passive L-matching network is slightly better than the design with the transmission lines due to the use of lossless components in the matching network. 
\begin{figure}
    \centering
\resizebox{0.5\textwidth}{0.5\textwidth}{

%\begin{circuitikz}%[american]
\ctikzset{bipoles/thickness=1}
%\begin{circuitikz}[line width=1pt]
\begin{circuitikz}[line width=0.75pt]

\draw(0,2) node[/tikz/circuitikz/bipoles/length=30pt,op amp, anchor=-] (opamp) {}
 (opamp.out) to [short] ++(0,1.5) to [/tikz/circuitikz/bipoles/length=30pt,R,l_=$R_f$]  ++(-1.8,0) to (opamp.-)
(opamp.+) node[ground]{};
\draw(0,-1) node[/tikz/circuitikz/bipoles/length=30pt,op amp, anchor=-] (opamp1) {}
 (opamp1.out) to [short] ++(0,1.5) to [/tikz/circuitikz/bipoles/length=30pt,R,l_=$R_f$]  ++(-1.8,0) to (opamp1.-)
 (opamp1.+) node[ground]{};
\draw(0,-4) node[/tikz/circuitikz/bipoles/length=30pt,op amp, anchor=-] (opamp2) {}
 (opamp2.out) to [short] ++(0,1.5) to [/tikz/circuitikz/bipoles/length=30pt,R,l_=$R_f$]  ++(-1.8,0) to (opamp2.-)
(opamp2.+) node[ground]{};
\draw(0,-7) node[/tikz/circuitikz/bipoles/length=30pt,op amp, anchor=-] (opamp3) {}
 (opamp3.out) to [short] ++(0,1.5) to [/tikz/circuitikz/bipoles/length=30pt,R,l_=$R_f$]  ++(-1.8,0) to (opamp3.-)
 (opamp3.+) node[ground]{};

\draw (opamp.out) node[right]{$Out_I^+$};
\draw (opamp1.out) node[right]{$Out_I^-$};
\draw (opamp2.out) node[right]{$Out_Q^+$};
\draw (opamp3.out) node[right]{$Out_Q^-$};

\draw (opamp.-) to [short]  ++(-2,0) node[nmos,anchor=S,rotate=-90,yscale=-1](nmos_0){};
\draw (opamp1.-) to [short] ++(-2,0) node[nmos,anchor=S,rotate=-90,yscale=-1](nmos_180){};

\draw (opamp2.-)  to [short]  ++(-2,0) node[nmos,anchor=S,rotate=-90,yscale=-1](nmos_90){};
\draw (opamp3.-)  to [short]  ++(-2,0) node[nmos,anchor=S,rotate=-90,yscale=-1](nmos_270){};

\draw (nmos_270.G) to [open] ++(0,-0.1) node[label=$270^\circ$]{} ;
\draw (nmos_180.G) to [open] ++(0,-0.1) node[label=$180^\circ$]{} ;
\draw (nmos_0.G) to [open] ++(0,-0.1) node[label=$0^\circ$]{} ;
\draw (nmos_90.G) to [open] ++(0,-0.1) node[label=$90^\circ$]{} ;

\draw (nmos_90.D) to [short] (nmos_270.D);
\draw (nmos_0.D) to [short] (nmos_180.D);

\draw (nmos_90.D)++(0,-1.5) to [/tikz/circuitikz/bipoles/length=30pt,L,l_=$L_{ser}$,mirror] node[](C1){} ++(-2,0) to [/tikz/circuitikz/bipoles/length=25pt,C,l=$C_{sh}$] ++(0,-1.5) node[ground]{};
\draw (nmos_0.D)++(0,-1.5) to [/tikz/circuitikz/bipoles/length=30pt,L,l_=$L_{ser}$,mirror] node[](C2){}  ++(-2,0) to [/tikz/circuitikz/bipoles/length=25pt,C,l=$C_{sh}$] ++(0,-1.5) node[ground]{};
\draw(nmos_90.D)++(-1.5,-1.5) to [short] ++(-1,0) to [short] ++(0,6) to [short] ++(1,0);
\draw (-6.05,-2.5) to [/tikz/circuitikz/bipoles/length=20pt,R,l_=$R_S$] ++(-1.5,0) to [/tikz/circuitikz/bipoles/length=30pt,american,sV,l_=$V_S$] ++(0,-1.5)  node[ground]{};

\draw(opamp.-)++(-1.5,0) to [/tikz/circuitikz/bipoles/length=20pt,C=$C_{BB}$] ++(0,-0.7) node[ground]{};
\draw(opamp1.-)++(-1.5,0) to [/tikz/circuitikz/bipoles/length=20pt,C=$C_{BB}$] ++(0,-0.7) node[ground]{};
\draw(opamp2.-)++(-1.5,0) to [/tikz/circuitikz/bipoles/length=20pt,C=$C_{BB}$] ++(0,-0.7) node[ground]{};
\draw(opamp3.-)++(-1.5,0) to [/tikz/circuitikz/bipoles/length=20pt,C=$C_{BB}$] ++(0,-0.7) node[ground]{};
\end{circuitikz} }

    \caption{Mixer-first Receiver with L-Matching Network}
    \label{fig:RX_MN1}
\end{figure}
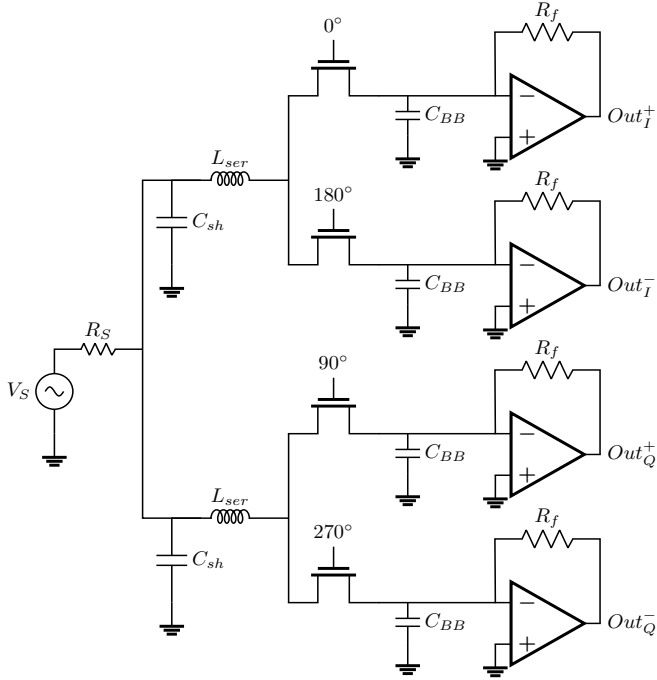
%In the design proposed in \cite{JSSC_sashank} and in section \ref{chap: intro}, matching was achieved using a $50\Omega$ physical resistor which added a 3dB penalty to the noise figure. 

\begin{figure}
    \centering
\begin{tikzpicture}
\begin{axis}[
    width=0.5\textwidth,
    height=0.35\textwidth,
    xlabel={Frequency [GHz]},
    ylabel={Noise Figure [dB]},
    xmin=10, xmax=50,
    ymin=3.5, ymax=22,
    xtick={10,15,20,25,30,35,40,45,50},
    ytick={4,6,8,10,12,14,16,18,20,22},
    legend pos=north west,
    grid=major,
    x label style={at={(axis description cs:0.5,0.0)}},
    y label style={at={(axis description cs:0.07,0.5)}},]

    \addplot[
    color=black,
    mark=square,
    mark size = {2},
    thick
    ]
    coordinates {
    (10,10)(15,10.54)(20,11.16)(25,11.81)(30,12.48)(35,13.14)(40,13.8)(45,14.43)(50,15)
    };
    
    \addlegendentry{Rx W/ 50$\Omega$ resistor}    

\addplot[
    color=blue,
    mark=o,
    mark size = {2},
    thick
    ]
    coordinates {
    (10,8.74)(15,7.3)(20,6.69)(25,6.31)(30,6.11)(35,6.81)(40,9.1)(45,12.36)(50,15.68)
    };
    \addlegendentry{Rx W/ Tlines}
    \addplot[
    color=red,
    mark=triangle,
    mark size = {2},
    thick
    ]
    coordinates {
    (10,8.84)(15,6.42)(20,5.05)(25,4.4)(30,4.15)(35,4.35)(40,5.37)(45,7.36)(50,12.57)
    };
    \addlegendentry{Rx W/ Lossless Matching Network}

    %\node (source) at (axis cs:30,4.25){};
    %\node (destination) at (axis cs:30,12.38){};
    %\draw [thick, <->] (source) -- (destination);
    %\node (label) at (axis cs:32,9.5){};
\end{axis}
\end{tikzpicture}
    \caption{Comparison of NF Across Three Different RX Designs}
    \label{fig:RX_MN}
\end{figure}
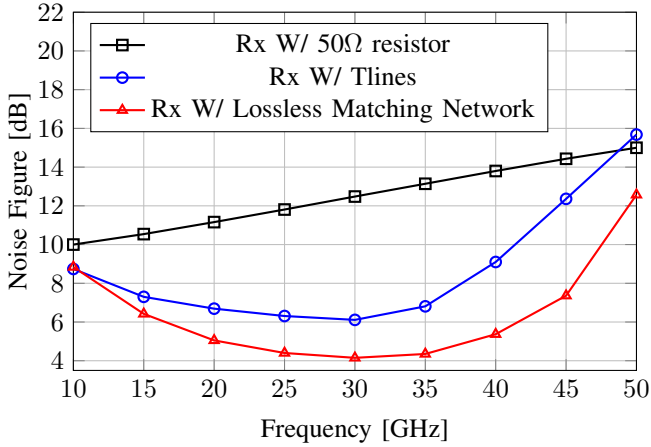

%\subsection{Limitations}

\begin{figure}

    \centering
     \resizebox{0.5\textwidth}{0.5\textwidth}{
\ctikzset{bipoles/thickness=1}
\begin{circuitikz}[line width=0.75pt]

\draw(0,2) node[/tikz/circuitikz/bipoles/length=30pt,op amp, anchor=-] (opamp) {}
 (opamp.out) to [short] ++(0,1.5) to [/tikz/circuitikz/bipoles/length=30pt,R,l_=$R_f$]  ++(-1.8,0) to (opamp.-)
(opamp.+) node[ground]{};
\draw(0,-1) node[/tikz/circuitikz/bipoles/length=30pt,op amp, anchor=-] (opamp1) {}
 (opamp1.out) to [short] ++(0,1.5) to [/tikz/circuitikz/bipoles/length=30pt,R,l_=$R_f$]  ++(-1.8,0) to (opamp1.-)
 (opamp1.+) node[ground]{};
\draw(0,-4) node[/tikz/circuitikz/bipoles/length=30pt,op amp, anchor=-] (opamp2) {}
 (opamp2.out) to [short] ++(0,1.5) to [/tikz/circuitikz/bipoles/length=30pt,R,l_=$R_f$]  ++(-1.8,0) to (opamp2.-)
(opamp2.+) node[ground]{};
\draw(0,-7) node[/tikz/circuitikz/bipoles/length=30pt,op amp, anchor=-] (opamp3) {}
 (opamp3.out) to [short] ++(0,1.5) to [/tikz/circuitikz/bipoles/length=30pt,R,l_=$R_f$]  ++(-1.8,0) to (opamp3.-)
 (opamp3.+) node[ground]{};

\draw (opamp.out) node[right]{$Out_I^+$};
\draw (opamp1.out) node[right]{$Out_I^-$};
\draw (opamp2.out) node[right]{$Out_Q^+$};
\draw (opamp3.out) node[right]{$Out_Q^-$};

\draw (opamp.-) to [short]  ++(-2,0) node[nmos,anchor=S,rotate=-90,yscale=-1](nmos_0){};
\draw (opamp1.-) to [short] ++(-2,0) node[nmos,anchor=S,rotate=-90,yscale=-1](nmos_180){};

\draw (opamp2.-)  to [short]  ++(-2,0) node[nmos,anchor=S,rotate=-90,yscale=-1](nmos_90){};
\draw (opamp3.-)  to [short]  ++(-2,0) node[nmos,anchor=S,rotate=-90,yscale=-1](nmos_270){};

\draw (nmos_270.G) to [open] ++(0,-0.1) node[label=$270^\circ$]{} ;
\draw (nmos_180.G) to [open] ++(0,-0.1) node[label=$180^\circ$]{} ;
\draw (nmos_0.G) to [open] ++(0,-0.1) node[label=$0^\circ$]{} ;
\draw (nmos_90.G) to [open] ++(0,-0.1) node[label=$90^\circ$]{} ;

\draw (nmos_90.D) to [short] (nmos_270.D);
\draw (nmos_0.D) to [short] (nmos_180.D);

\draw (nmos_90.D)++(0,-1.5) to [/tikz/circuitikz/bipoles/length=25pt,vC,l_=$C_{ser}$,mirror,invert] ++(-1.5,0) to [/tikz/circuitikz/bipoles/length=25pt,L,l_=$L_{ser}$,mirror]  ++(-1.5,0) to [/tikz/circuitikz/bipoles/length=25pt,vC=$C_{sh}$,invert] ++(0,-1.5) node[ground]{};
\draw (nmos_0.D)++(0,-1.5) to  [/tikz/circuitikz/bipoles/length=25pt,vC,l_=$C_{ser}$,mirror,invert] ++(-1.5,0) to [/tikz/circuitikz/bipoles/length=25pt,L,l_=$L_{ser}$,mirror] node[](C2){}  ++(-1.5,0) to [/tikz/circuitikz/bipoles/length=25pt,vC=$C_{sh}$,invert] ++(0,-1.5) node[ground]{};
\draw(nmos_90.D)++(-2.5,-1.5) to [short] ++(-1,0) to [short] ++(0,6) to [short] ++(1,0);
\draw (-7.05,-2.5) to [/tikz/circuitikz/bipoles/length=20pt,R,l_=$R_S$] ++(-1.5,0) to [/tikz/circuitikz/bipoles/length=30pt,american,sV,l_=$V_S$] ++(0,-1.5)  node[ground]{};

\draw(opamp.-)++(-1.5,0) to [/tikz/circuitikz/bipoles/length=20pt,C=$C_{BB}$] ++(0,-0.7) node[ground]{};
\draw(opamp1.-)++(-1.5,0) to [/tikz/circuitikz/bipoles/length=20pt,C=$C_{BB}$] ++(0,-0.7) node[ground]{};
\draw(opamp2.-)++(-1.5,0) to [/tikz/circuitikz/bipoles/length=20pt,C=$C_{BB}$] ++(0,-0.7) node[ground]{};
\draw(opamp3.-)++(-1.5,0) to [/tikz/circuitikz/bipoles/length=20pt,C=$C_{BB}$] ++(0,-0.7) node[ground]{};
\end{circuitikz} }
    \caption{Mixer-first Receiver with Tunable Matching Networks}
    \label{fig:Rx_TunableMN}
\end{figure}
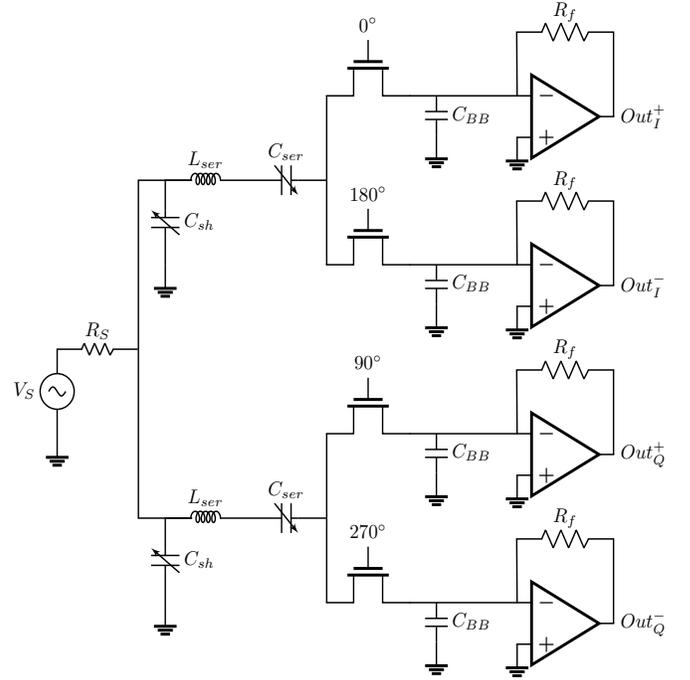
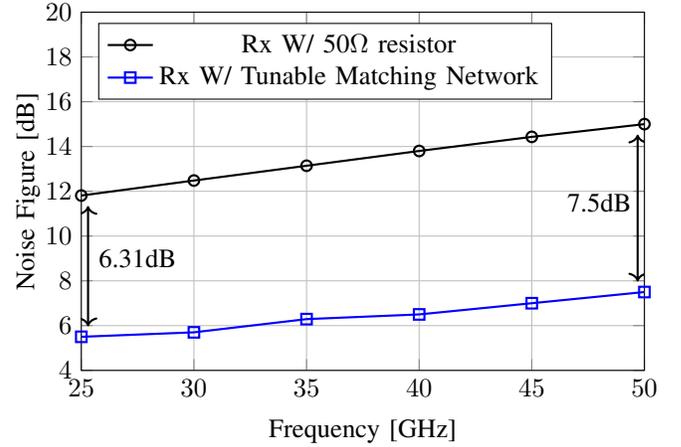
\begin{figure}
    \centering

\begin{tikzpicture}
\begin{axis}[
    width=0.5\textwidth,
    height=0.35\textwidth,
    xlabel={Frequency [GHz]},
    ylabel={Noise Figure [dB]},
    xmin=25, xmax=50,
    ymin=4, ymax=20,
    xtick={25,30,35,40,45,50},
    ytick={2,4,6,8,10,12,14,16,18,20,22},
    legend pos=north west,
    ymajorgrids=true,
    xmajorgrids=true,
    x label style={at={(axis description cs:0.5,0.0)}},
    y label style={at={(axis description cs:0.07,0.5)}},]
    \addplot[
    color=black,
    mark=o,
    mark size = {2},
    thick
    ]
    coordinates {
    (25,11.81)(30,12.48)(35,13.14)(40,13.8)(45,14.43)(50,15)
    };
    
    \addlegendentry{Rx W/ 50$\Omega$ resistor}
\addplot[
    color=blue,
    mark=square,
    mark size = {2},
    thick
    ]
    coordinates {
    (25,5.5)(30,5.7)(35,6.29)(40,6.5)(45,7)(50,7.5)
    };
    \addlegendentry{Rx W/ Tunable Matching Network}
    \node (source) at (axis cs:25.3,5.55){};
    \node (destination) at (axis cs:25.3,11.76){};
    \draw [thick, <->] (source) -- (destination);
    \node (label) at (axis cs:27.5,9){6.31dB};
    
    \node (source) at (axis cs:49.7,7.55){};
    \node (destination) at (axis cs:49.7,14.95){};
    \draw [thick, <->] (source) -- (destination);
    \node (label) at (axis cs:48,11.5){7.5dB};

\end{axis}
\end{tikzpicture}
    \caption{Noise figure values of the proposed linear receiver compared to the design proposed in the previous section. Ideal LO, baseband amplifier, and $R_{sw}=6\Omega$}
    \label{fig:my_label}
\end{figure}
\subsection{Tunable Matching Network}
Using passive components in the matching network will enable the possibility of using the receiver in wideband applications. To design a wideband receiver, the matching network needs to be tunable, meaning that both the inductor and the capacitor need to be adjustable. Fig. \ref{fig:Rx_TunableMN} shows the schematic of the receiver with the tunable matching network. The variable capacitor can be implemented with either a varactor or a capacitor bank. The tunability of the inductor is more difficult to realize since inductors occupy larger amounts of area and implementing an inductor bank would be area consuming. Hence, the variable inductor needs to be implemented using a different approach. In this design, the tunable inductor is implemented with an inductor in series with a capacitor bank. The series impedance of that combination can be written as
\begin{equation} \label{eq:1}
Z_{series}=j\omega L_{ser}+\frac{1}{j\omega C_{ser}}=j(\omega L_{ser}- \frac{1}{\omega C_{ser}})
\end{equation}
By looking at equation \ref{eq:1}, it can be shown that changing the value of the series capacitor has the same effect as implementing a tunable inductor. Decreasing the value of the series capacitor while keeping the inductor fixed will be the equivalent to decreasing the value of the inductor.
 %Figure (include figure shows the movement on the smith chart of the three element. 

The shunt capacitor can be implemented with a varactor since its values vary slightly over different frequencies. The series capacitor, on the other hand, needs to be implemented with a capacitor bank. 
\subsection{Linearity}
The linearity of the receiver is governed by linearity of the mixers and the baseband amplifier. This can be seen by analyzing the overall IIP3 of the receiver which is given with the following equation:
\begin{equation}
    \frac{1}{V_{IIP3}^2}=\frac{a_{MN}^2}{V_{IIP3,mixer}^2}+\frac{a_{MN}^2 a_{mixers}^2}{V_{IIP3,BB}^2}
\end{equation}

The gain of a 4-phase mixers, $a_{mixers}$, is approximately 1. The voltage gain of the matching network, $a_{MN}$, is less than unity, which means that the IIP3 of the mixers is effectively increased. The linearity limit imposed by the baseband amplifier is mitigated by using feedback linearization, which is a technique previously proposed in \cite{JSSC_sashank}. By choosing a large open loop gain of the baseband amplifier, the input of the baseband amplifier can be nearly a virtual ground. This can be visualized in Fig. \ref{fig:swLinearity}, where $V_x$ would be almost zero by design. Making the input of the baseband amplifier's a virtual ground means that input swing to the baseband amplifier is minimized which limits the effect of the baseband non-linearity on the receiver's overall linearity. 

As for the mixer's linearity, the large gain of the baseband amplifier means that source terminal of the mixer, which is the input to the baseband amplifier, is a virtual ground. This will reduce the $V_{gs}$ swing of the mixers making them  more linear. This concept is similar to "bottom-plate" mixers discussed in \cite{bottom_plate}.

The linearity limit imposed by the mixers' $V_{ds}$ swing is mitigated by stepping down the input voltage with the matching network and using a smaller on-resistance for the mixers. Fig. \ref{fig:swLinearity} shows a simplified schematic of the receiver and the effect of using the matching network on $V_{ds}$ of the mixers. Assuming that the matching network is ideal, the output power of both matching networks will be
\begin{equation}
    P_{out}=\frac{P_{in}}{2}
    \label{eq:PoutPin}
\end{equation}
The power at the input, $P_{in}$, can be expressed as 
\begin{equation}
    P_{in}=\frac{V_{in}^2}{2R_s}
    \label{eq:pin}
\end{equation}
And the output power of the matching network can be expressed as
\begin{equation}
    P_{out}=\frac{V_{ds}^2}{2R_{sw}}
    \label{eq:pout}
\end{equation}
Solving for $V_{ds}$ in terms of $V_{in}$ by plugging \ref{eq:pout} and \ref{eq:pin} into \ref{eq:PoutPin}:
\begin{equation}
    V_{ds}=\frac{v_s}{2}\sqrt{\frac{R_{sw}}{2R_s}}
\end{equation}
With $R_{sw}=12\Omega$ and $R_s=50\Omega$:

\begin{equation}
    V_{ds}=\frac{v_s}{2}\sqrt{\frac{R_{sw}}{2R_s}}=\frac{v_s}{2}\sqrt{\frac{12}{100}}=0.17V_s
\end{equation}

The expression of $V_{ds}$ shows that for smaller $R_{sw}$, the drain source voltage swing will decrease making the receiver more linear. A smaller $V_{ds}$ results in a more linear mixer. The reduction of the $V_{ds}$ swing across the mixer would not be the limiting factor in the receiver linearity. 
\begin{figure}
    \centering
    \resizebox{0.45\textwidth}{0.5\textwidth}{

    \begin{circuitikz}[american]    
    \draw (8,-1.5) node[/tikz/circuitikz/bipoles/length=30pt, op amp,anchor=-](opamp3){};    
    \draw (8,-4.5) node[/tikz/circuitikz/bipoles/length=30pt, op amp,anchor=-](opamp4){} ;  
    \draw (opamp4.-)++(-0.5,-.1) to [open,v=\tiny$V_x$] ++(0,-0.8);   
    \draw (8,4.5) node[/tikz/circuitikz/bipoles/length=30pt, op amp,anchor=-](opamp1){};  
    \draw (8,1.5) node[/tikz/circuitikz/bipoles/length=30pt, op amp,anchor=-](opamp2){};
    \draw(1,0) to [short] ++(0,3) to [twoport,t=$8:1$,l=MN] ++(2.5,0) to [short] ++(0,-1.5) to [short] ++(0.4,0) to [short,*-*] ++(0.6,0) to [/tikz/circuitikz/bipoles/length=20pt,R=$R_{sw}$] ++(1.5,0) to [short] (opamp2.-);
    \draw(3.9,4.5) to [short] ++(0.55,0.15);
    \draw(3.5,3) to [short] ++(0,1.5) to [short] ++(0.4,0) to [open, *-*] ++(0.6,0) to [/tikz/circuitikz/bipoles/length=20pt,R=$R_{sw}$] ++(1.5,0) to [short] (opamp1.-);
    \draw(3.9,-1.5) to [short] ++(0.55,0.15);
    \draw(1,0) to [short] ++(0,-3) to [twoport,t=$8:1$,l=MN] ++(2.5,0) to [short] ++(0,-1.5) to [short] ++(0.4,0) to [short,*-*] ++(0.6,0) to [/tikz/circuitikz/bipoles/length=20pt,R=$R_{sw}$,v=$V_{ds}$] ++(1.5,0) to [short] (opamp4.-) ;
    \draw(3.5,-3) to [short] ++(0,1.5) to [short] ++(0.4,0) to [open, *-*] ++(0.6,0) to [/tikz/circuitikz/bipoles/length=20pt,R=$R_{sw}$]  ++(1.5,0) to [short] (opamp3.-); 
  
    \draw (1,0) to [/tikz/circuitikz/bipoles/length=20pt,R,l_=$R_s$] ++(-1.5,0) to [/tikz/circuitikz/bipoles/length=30pt,sV,l_=$V_s$] ++(0,-1) node[ground]{};  
    \draw(opamp1.-)++(-1,0) to [/tikz/circuitikz/bipoles/length=20pt,C,l_=\tiny$C_{BB}$] ++(0,-0.7) node[ground]{};
    \draw(opamp2.-)++(-1,0) to [/tikz/circuitikz/bipoles/length=20pt,C,l_=\tiny$C_{BB}$] ++(0,-0.7) node[ground]{};
    \draw(opamp3.-)++(-1,0) to [/tikz/circuitikz/bipoles/length=20pt,C,l_=\tiny$C_{BB}$] ++(0,-0.7) node[ground]{};
    \draw(opamp4.-)++(-1,0) to [/tikz/circuitikz/bipoles/length=20pt,C,l_=\tiny$C_{BB}$] ++(0,-0.7) node[ground]{};
    %feedback resistors 
    \draw (opamp1.out) to [short] ++(0,1) to [/tikz/circuitikz/bipoles/length=20pt,R,l_=$R_f$] ++(-1.8,0) to [short] (opamp1.-);
    \draw (opamp2.out) to [short] ++(0,1) to [/tikz/circuitikz/bipoles/length=20pt,R,l_=$R_f$] ++(-1.8,0) to [short] (opamp2.-);
    \draw (opamp3.out) to [short] ++(0,1) to [/tikz/circuitikz/bipoles/length=20pt,R,l_=$R_f$] ++(-1.8,0) to [short] (opamp3.-);
    \draw (opamp4.out) to [short] ++(0,1) to [/tikz/circuitikz/bipoles/length=20pt,R,l_=$R_f$] ++(-1.8,0) to [short] (opamp4.-);
    \draw (opamp1.+) node[ground]{};
    \draw (opamp2.+) node[ground]{};
    \draw (opamp3.+) node[ground]{};
    \draw (opamp4.+) node[ground]{};
    \draw(1,-3) to [short] ++(-0,-0.5) to [open,*-*,v=$V_{in}$] ++(0,-1.5) node[ground]{};
    
    \end{circuitikz}}
    \caption{Simplified schematics of figure \ref{fig:Rx_TunableMN}}
    \label{fig:swLinearity}
\end{figure}

\begin{figure}
    \centering
     \resizebox{0.45\textwidth}{0.13\textwidth}{

    \begin{circuitikz}[american]
    \draw (16,0) node[op amp,anchor=-,t=A](opamp5){};
    \draw (opamp5.out) to [short] ++(0,1.5) to [/tikz/circuitikz/bipoles/length=20pt,R,l_=$\gamma R_f$] ++(-2.4,0) to [short] (opamp5.-);
    \draw (opamp5.+) node[ground]{};
    \draw (opamp5.-)++(-0.75,0) to [/tikz/circuitikz/bipoles/length=25pt,R,l_=$R_{sh}$] ++(0,-1) node[ground]{};
    \draw (opamp5.-)++(-2,0) to [/tikz/circuitikz/bipoles/length=25pt,R,l_=$R_{OL}$] ++(0,-1) node[ground]{};
    \draw(opamp5.-) to [short] ++(-3,0) to [/tikz/circuitikz/bipoles/length=25pt,R,l_=$R_{sw}$] ++(-2,0) to [twoport,t=$4:1$,l_=MN] ++(-1.5,0) to [/tikz/circuitikz/bipoles/length=25pt,R,l_=$R_s$] ++(-1.5,0) to [/tikz/circuitikz/bipoles/length=30pt,sV,l_=$V_s$] ++(0,-1) node[ground]{}; 
    \end{circuitikz}}
    
    \caption{LTI equivalent circuits for figures \ref{fig:Rx_TunableMN}  and \ref{fig:swLinearity}. Where $R_{sh}$ is the re-radiation resistance and $R_{OL}$ is the overlap resistance. The two 8:1 matching networks in figure \ref{fig:swLinearity} can be replaced by one 4:1 matching network in the LTI model}
    \label{fig:LTImodel}
\end{figure}
\subsection{Noise Figure}
As discussed in section \ref{chap: intro}, the noise figure of the design proposed in \cite{JSSC_sashank} is high due to the use of a 50$\Omega$ resistor and the charge sharing current caused by the overlapping LO waveform. In addition to eliminating the physical resistor from the design, this design help improve the noise figure in different ways. 

Fig. \ref{fig:LTImodel} shows the linear time invariant model of the proposed design. The noise figure of the design can be expressed with the following equation:
\begin{equation}
\begin{split}
    F=1+\frac{R_{sw}}{R_s}+\frac{R_{sh}}{R_s}\left(\frac{R_{sw}+R_s}{R_{sh}}\right)^2 +\\\frac{R_{OL}}{R_s}\left(\frac{R_{sw}+R_s}{R_{OL}}\right)^2
    \end{split}
    \label{eq:NF}
\end{equation}
The noise of the baseband amplifier is neglected for simplicity and the matching network is assumed to be ideal.
\subsubsection{LO Harmonic Suppression}
The proposed architecture offers LO harmonic filtering. The presence of high frequency LO harmonics means that interferes at those frequencies will be down converted to baseband, degrading the signal-to-noise ratio at the output and the receiver’s noise figure. The matching network between the antenna and the mixer acts as a low-pass filter. Interferes at the LO harmonics will be attenuated before getting down converted to baseband. This will improve the signal-to-noise ratio at the output and will improve the noise figure as opposed to the design in \cite{JSSC_sashank}. The filtering effect makes this receiver architecture a harmonic rejection receiver. Unlike conventional harmonic rejection receivers, the tunable nature of the matching network allows for wideband operation.  

In addition to being a harmonic rejection receiver, the matching network works as a filter for the current flowing from the mixers to the antenna. In equation \ref{eq:NF}, \cite{Rsh} showed that $R_{sh}$ is due to the power losses resulting from the up-conversion of the LO harmonics to the antenna. $R_{sh}$ can be expressed as the parallel combination of the antenna's impedance at the odd harmonics of the LO. Since the matching network is a low pass filter, the impedance looking at the antenna from the mixers side would be large.   

The large value of $R_{sh}$ would translate to less loss in the receiver and that would improve the noise figure. Equation \ref{eq:NF} shows how with increasing $R_{sh}$, the term where $R_{sh}$ appears would tend to zero with increasing $R_{sh}$

\subsubsection{Charge Sharing}
In addition to being part of the matching network, the series inductor in the matching network offers another benefit. 
the inductor provides filtering to the charge sharing current the I and Q paths. This will consequently improve the noise figure of the receiver. The use of an inductor is also reported in \cite{Molnar_indu} to reduce charge sharing. 

In equation (include equation), $R_{OL}$ is used to model the effect of the overlapping LO waveform. $R_{OL}$ is proportional to the resistance per path of the N-path filter and inversely proportional to the overlap time between the LO waveform. $R_{OL}$ is infinite when 25\% duty-cycle LO waveform is used.
$R_{OL}$ can be approximated using the following equation:
\begin{equation}
    R_{OL}\propto\frac{R_{path}}{\omega_{LO}\tau_{overlap}}
\end{equation}
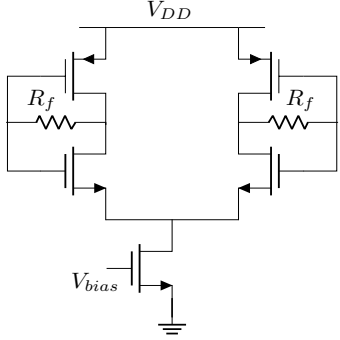
\begin{figure}
    \centering
        \resizebox{0.25\textwidth}{0.25\textwidth}{

\begin{circuitikz}[american]  
\ctikzset{transistors/arrow pos=end}
\ctikzset{tripoles/mos style/arrows}
\draw (0,0) node[nmos,anchor=S,xscale=-1](nmos_0){};
\draw (0,1.5) node[pmos,anchor=D,xscale=-1](pmos_0){};
\draw (nmos_0.D) to (pmos_0.D);
\draw (nmos_0.G) to [short] ++(0.5,0) to [short] ++(0,1.5) to (pmos_0.G);
\draw (-2,0) node[nmos,anchor=S](nmos_1){};
\draw (-2,1.5) node[pmos,anchor=D](pmos_1){};
\draw (nmos_1.D) to (pmos_1.D);
\draw (nmos_1.G) to [short] ++(-0.5,0) to [short] ++(0,1.5) to (pmos_1.G);
\draw (nmos_0.S) to (nmos_1.S);
\draw (nmos_0.D) to [/tikz/circuitikz/bipoles/length=20pt,R] ++(1.5,0);
\draw (nmos_1.D) to [/tikz/circuitikz/bipoles/length=20pt,R] ++(-1.5,0);
\draw (-1,0) node[nmos,anchor=D](bias){}
(bias.S) to ++(0,0.3) node[ground]{};
\draw (pmos_0.S) to (pmos_1.S) to [short] ++(-0.4,0);
\draw (pmos_0.S) to [short] ++(0.4,0);
\draw (-0.4,3.3) to [open, l=$V_{DD}$] ++(0,0);
\draw (-1.5,-1) to [open,l=$V_{bias}$] ++(0,0);
\draw (1.45,1.9) to [open,l=$R_f$] ++(0,0);
\draw (-2.45,1.9) to [open,l=$R_f$] ++(0,0);

    \end{circuitikz}}
    \caption{Inverter-based Baseband amplifier}
    \label{fig:BBAmp}
\end{figure}

\begin{figure}
    \centering
        \resizebox{0.5\textwidth}{0.15\textwidth}{

    \includegraphics{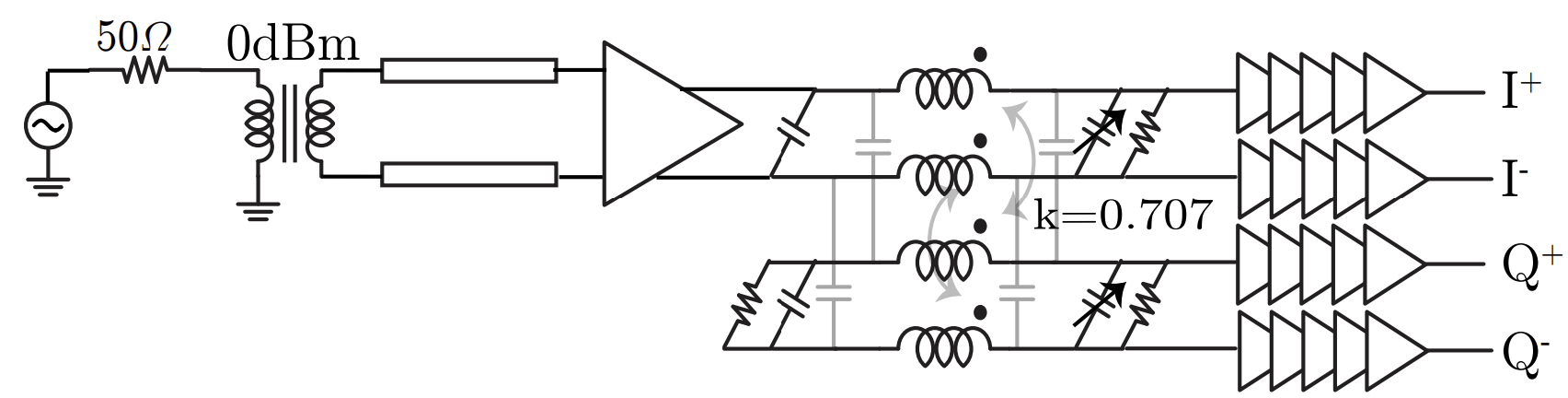}}
    \caption{Schematic of the LO Chain \cite{Krishnamurthy:EECS-2021-231}}
    \label{fig:LOChain}
\end{figure}

\section{Circuit Implementation}
\label{chap:4}
\subsection{Baseband Amplifier}
The baseband amplifier is implemented with and inverter-based amplifier shown in figure \ref{fig:BBAmp}. Open loop gain of the amplifier is 36dB and the feedback resistor is programmable. 
\subsection{LO Generation}

The LO chain used is shown in figure \ref{fig:LOChain}. The single-ended LO input is converted to a differential waveform using a balun. Using inverter-based buffers, the differential waveform is fed into a quadrature hybrid to generate the four-phase 50\% LO waveforms. Inverter-based LO buffers are used after the quadrature hybrid to drive the mixers. This LO chain used is similar to the one used in \cite{JSSC_sashank}.

\subsection{Mixers}
NMOS transistors are used for the mixers. The choice of the mixers' size affects the performance of the receiver. Choosing a large device would make the on-resistance of the mixer smaller, improving the noise figure. The simulation results showed a 0.3dB improvement in the noise figure when using a switch with on-resistance 6$\Omega$ instead of 12$\Omega$. The mixers' are driven by an ideal voltage source with overlapping waveforms. 

The larger transistor size means that the gate capacitance of the mixers will be larger, making it more difficult to drive the mixers with a reasonable power consumption. The larger mixers also mean adding more parasitic capacitance to the input of the mixer, leading to more loss in the signal path and further degradation in the noise figure. Hence, the final design uses switches with $W/L=\SI{27}{\micro\meter}/\SI{30}{\nano\meter}$ instead of a larger device size, with on-resistance of 12$\Omega$.

%Because of the reasons discussed in this section, the final design uses switches with $W/L=\SI{27}{\micro\meter}/\SI{30}{\nano\meter}$ instead of a larger device size. And the on-resistance of the switches are 12$\Omega$.

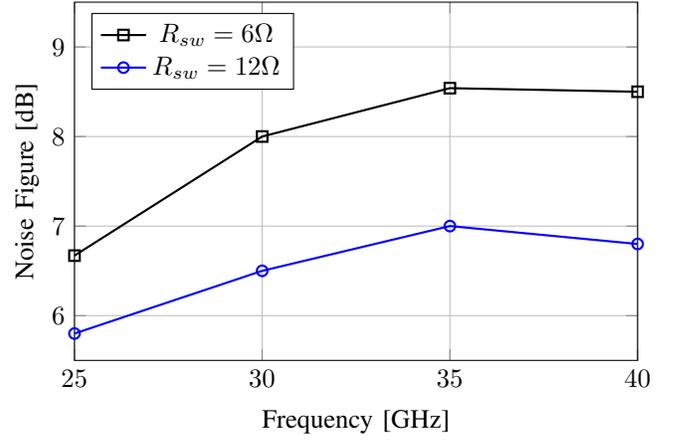
\begin{figure}
    \centering
\begin{tikzpicture}
\begin{axis}[
    width=0.5\textwidth,
    height=0.35\textwidth,
    xlabel={Frequency [GHz]},
    ylabel={Noise Figure [dB]},
    xmin=25, xmax=40,
    ymin=5.5, ymax=9.5,
    xtick={25,30,35,40},
    ytick={5,6,7,8,9,10},
    legend pos=north west,
    grid=both,
    y label style={at={(axis description cs:0.08,0.5)}},]    
    \addplot[
    color=black, mark=square, mark size = {2}, thick,
    ]
    coordinates {
    (25,6.67)(30,8)(35,8.54)(40,8.5)
    };
    
    \addlegendentry{$R_{sw}=6\Omega$}
\addplot[
    color=blue, mark=o, mark size = {2}, thick,
    ]
    coordinates {
    (25,5.8)(30,6.5)(35,7)(40,6.8)
    };
    \addlegendentry{$R_{sw}=12\Omega$}
    
\end{axis}
\end{tikzpicture}

    \caption{\centering Noise Figure of Rx with Fully Tunable Lossy Matching Network, Transistor switches, and LO Chain}
    \label{fig:NF_LO}
\end{figure}
\subsection{Matching Network} 
\begin{figure}
    \centering     
    \resizebox{0.45\textwidth}{0.35\textwidth}{

    \includegraphics{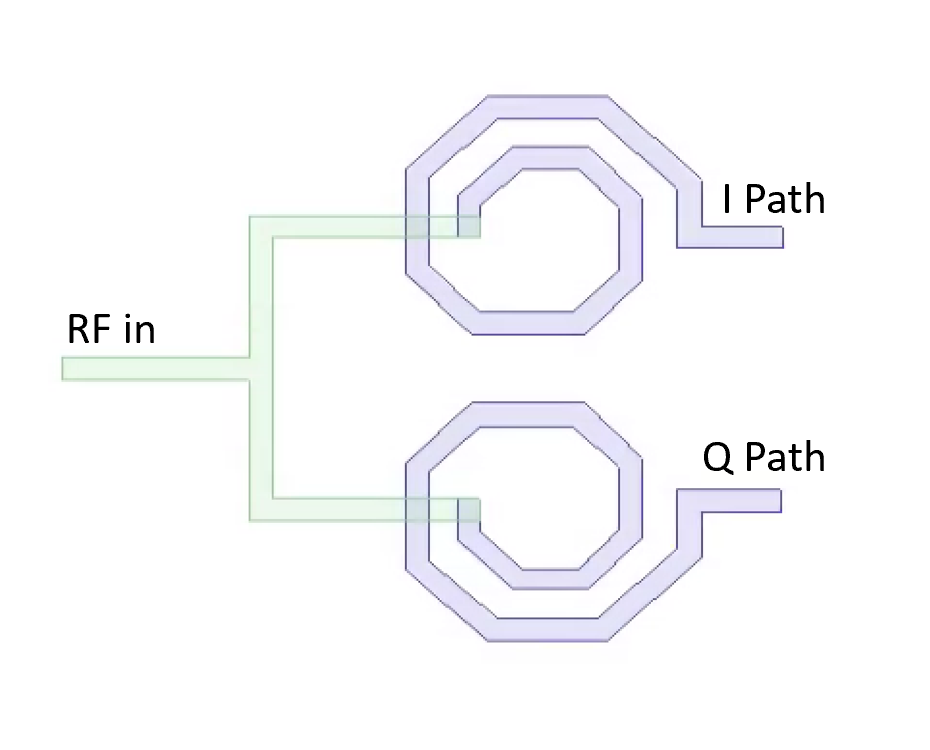}}
    \caption{Inductors' Layout}
    \label{fig:IndLayout}
\end{figure}

The pad and the ESD capacitance at the input can be included to implement the shunt capacitor. The inductors are implemented with octagonal single-turn inductor to maximize the quality factor of the inductor. Fig. \ref{fig:IndLayout} shows the layout of the inductors. The current in the inner most branch are in the same direction, which means that mutual inductance will increase the inductance of both inductors. The series capacitor is implemented with a capacitor bank.
\newcolumntype{P}[1]{>{\centering\arraybackslash}p{#1}}
\newcolumntype{M}[1]{>{\centering\arraybackslash}m{#1}}
\begin{table*}[!t]
\centering
\begin{threeparttable}
\renewcommand{\arraystretch}{1.3}
    \caption{Comparison with mixer-first receivers greater than 25GHz}
 
    \begin{tabular}{|M{1.7cm}|M{1.6cm}|M{1.8cm}|M{1.9cm}|M{1.6cm}|M{1.6cm}|M{1.6cm}|M{2.3cm}|}
        \hline
                               &  \textbf{Moroni} \cite{moroni2012broadband} RFIC 2012  & \textbf{Wilson} \cite{wilson201620} RFIC 2016 & \textbf{\small Krishnamurthy} \cite{krishnamurthyenhanced} RFIC2019 & \textbf{Iotti} \cite{iottimixerfirst} JSSC2020 &  \textbf{Ahmed} \cite{cicc2020} CICC2020  &  \textbf{This work} \\\hline
Technology & 65nm CMOS & 45nm SOI  & 28nm CMOS & 28nm CMOS& 22nm FD-SOI &  \textbf{28nm CMOS} \\\hline                          
$f_{RF}$ (GHz) & 49 -- 67 & 20--30 & 10 -- 35 & 70 -- 100  & 43 -- 97 & \textbf{25 -- 40} \\\hline
Voltage gain (dB) & 13 & 8 -- 20.6  & 11.5 -- 14.5 & 19.5 -- 25.3 & 12 -- 15 & \textbf{18} $^\dagger$\\\hline
%IP1dB (dBm) & -12 & -9.3 -- -13 & -16.8 -- -24 & -45  & -5.6 -- -8 & \textbf{-2 -- 0} $^\dagger$ \\\hline
%Best case OP1dBV & -10 & -3.4 & -8.3 & -11 & -4 & \textbf{1.5} \\\hline
In-band IIP3 (dBm) & - & -2.3 -- -9.7  & +10 -- +14.1 & - & 0 -- +4 & \textbf{+3.6 -- +4.2} $^\dagger$ \\\hline
NF (dB) & 11--14 & 8 & 12.5 -- 19.2  & 8 -- 12.7 & 12.5 -- 16.5 & \textbf{6.8 -- 7.5} $^\sharp$ \\\hline
DC power (mW) & 14 & 41 (at 24GHz)  & 22.8 (Baseband);     19 -- 37 (LO)  & 12 & 36 & \textbf{22.8} (Baseband);     \textbf{19 -- 37} (LO) \\\hline
Supply (V) & 1.2 & 0.9/1.8 & 1.2 & 1  & - & \textbf{1.2} \\\hline
\end{tabular}
%\footnotetext[5]{Estimated from figure.}
\begin{tablenotes}

\item $^\dagger$ Measurements reported at nominal setting ($R_F=1\mathrm{k\Omega}$), across $f_{LO}$.
\item $^\sharp$ NF varies from 6.8 -- 7.5 dB for $f_{LO} = 25 - 40 \mathrm{GHz}$.
\end{tablenotes}
\label{table1}  
\end{threeparttable}
\end{table*}

%\section{Inductors} 

\begin{figure}
    \centering
\begin{tikzpicture}
\begin{axis}[
    axis y line*=right,
    width=0.5\textwidth,
    height=0.35\textwidth,
    xlabel={Frequency [GHz]},
    ylabel={IIP3 [dBm]},
    xmin=25, xmax=40,
    ymin=2, ymax=8,
    xtick={25,30,35,40},
    ytick={2,4,6,8},
    legend style={at={(0.95,0.95)},legend columns=-1},
    ymajorgrids=true,
    xmajorgrids=true,
    y label style={at={(axis description cs:1.23,0.5)}},
]

\addplot[
    color=blue,
    mark=o,
    mark size = {2},
    thick
    ]
    coordinates {
    (25,4.14)(30,3.67)(35,3.4)(40,4.7)
    }; \label{plot_one}
    \end{axis}
\begin{axis}[
    axis y line*=left,
    width=0.5\textwidth,
    height=0.35\textwidth,
    xlabel={Frequency [GHz]},
    ylabel={Conversion Gain and NF [dB]},
    xmin=25, xmax=40,
    ymin=6, ymax=24,
    xtick={25,30,35,40},
    ytick={6,8,10,12,14,16,18,20,22,24},
    legend style={at={(0.95,0.95)},legend columns=-1},
    ymajorgrids=true,
    xmajorgrids=true,
    y label style={at={(axis description cs:0.07,0.5)}},
]\addlegendimage{/pgfplots/refstyle=plot_one}\addlegendentry{plot 1}
    \addplot[
    color=black,
    mark=square,
    mark size = {2},
    thick
    ]
    coordinates {
    (25,18.9)(30,19.1)(35,19)(40,18.5)
    };

    \addplot[
    color=red,
    mark=triangle,
    mark size = {3},
    thick
    ]
    coordinates {
    (25,7.5)(30,7.2)(35,6.9)(40,7.2)
    };
   % \addlegendentry{Noise Figure [dB]}
    \legend{IIP3 [dBm],Gain [dB], NF [dB]}
\end{axis}    

\end{tikzpicture}
    \caption{Post-layout Simulations of the design}
    \label{fig:finalsims}
\end{figure}
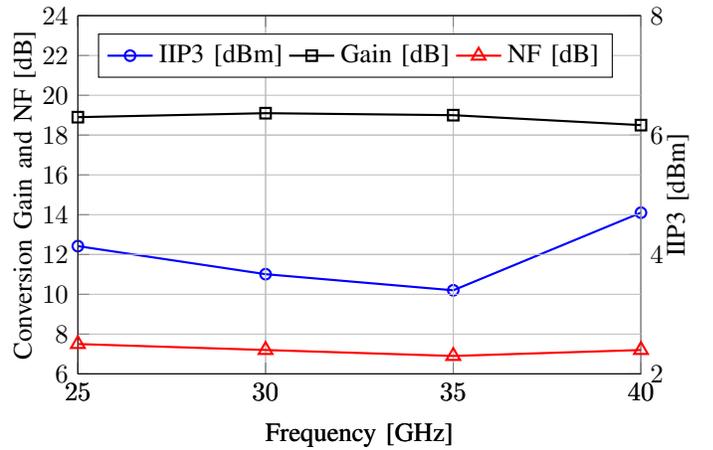

Fig. \ref{fig:finalsims} shows the final results of the post-layout simulated design. $R_f=\SI{1}{\kilo\Omega}$ is used. Noise analysis in virtuoso shows that the noise of $R_f$ contribute 30\% of the overall receiver noise. Choosing a larger value will improve the noise figure but will degrade the linearity of the receiver. The well known discontinuities in the BSIM4 \cite{BSIM} models made it difficult to simulate IP3 of the receiver. The source of the discontinuities are transistors in deep triode region, which in this case are the mixers. The mixers were replaced by ideal switches to simulate the IP3.

\section{Measurement}
\begin{figure}
        \centering
    \resizebox{0.35\textwidth}{0.3\textwidth}{
    \includegraphics[angle=0]{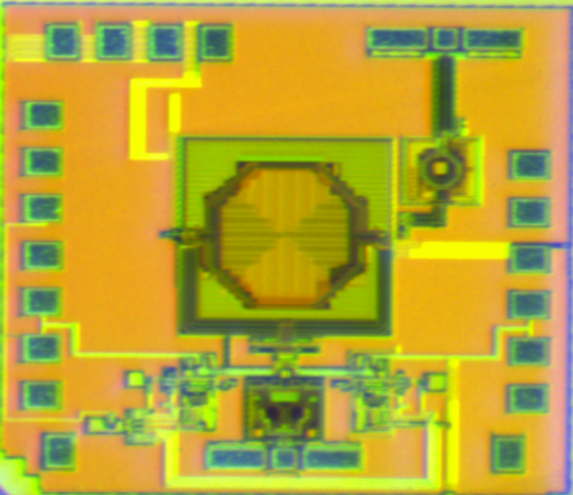}}
    \caption{Chip Micrograph in \SI{28}{\nano\meter} bulk CMOS process}
    \label{fig:ugraph}
\end{figure}

The design was taped out in \SI{28}{\nano\meter} bulk CMOS process and is awaiting measurement. Fig. \ref{fig:ugraph} show a micrograph of the test chip. The LO and the RF input are to be probed with GSG probes.  
%\IEEEPARstart{T}{his} demo file is intended to serve as a ``starter file''
%for IEEE journal papers produced under \LaTeX\ using
%IEEEtran.cls version 1.8b and later.

 
%\hfill November 7, 2022

\section{Conclusion}
\label{chap:5}
This paper proposed multiple designs for linear mixer-first receivers. Post-layout simulations showed an improvement in noise figure values compared to the design proposed in \cite{JSSC_sashank}. The final design achieved a noise figure of less than 8dB from a frequency range of 25GHz to 40GHz, and an IIP3 of +3.6dBm to +4.2dBm across the frequency range. Table \ref{table1} shows a comparison between the proposed design and the state-of-the-art mixer-first receiver greater than 25GHz. The noise figure value are lower than the other reported noise figure values. Although the IIP3 numbers are lower compared to the values reported in \cite{JSSC_sashank}, the values are improved compared to the other designs listed. The in-band IP3 values of the receiver can be improved by using a baseband amplifier with a large open loop gain. The larger amplifier gain would minimize the swing at the input of the baseband amplifier, which will consequently improve the linearity of the receiver. Improving the LO Chain will deliver a better LO waveform which would help with the linearity of the switch and the overall performance of the receiver. 

Switching to a different process will help with the performance limitation associated with using CMOS process. Using Fully Depleted Silicon-On-Insulator process will help with the loss in the signal and will result in passives with better performance. The use of FD-SOI is also beneficial to delivering square LO Waveform which would help with the performance of the receiver.

%Add a table of comparison

% or


%\appendices

%\section{}

\section*{Acknowledgment}

The authors would like to thank TSMC shuttle program for the chip fabrication and Hesham Beshary for his help with the tape-out.

\ifCLASSOPTIONcaptionsoff
  \newpage
\fi



\bibliographystyle{IEEEtran}
\bibliography{ref}
%\begin{thebibliography}{1}


%\end{thebibliography}
%\bibliographystyle{IEEEtran}
%\bibliography{ref}



%\newpage


%\vfill


\end{document}